\documentclass[]{article}

\usepackage{natbib}
\usepackage{amssymb}
\usepackage{amsmath}

\title{N-body Simulations}
\author{Michele Trenti \& Piet Hut}
\begin{document}

\begin{center}
{\huge \bf {Gravitational N-body Simulations}
}\vspace{1cm}

{\large Michele Trenti\footnote{trenti@stsci.edu}} 

{\small Space Telescope Science Institute, Baltimore, MD, 21210, U.S.}

\vspace{0.3cm}

{\small AND}

\vspace{0.3cm}

{\large Piet Hut} 

{\small Institute for Advanced Study, Princeton, NJ, 08540, U.S.}

\vspace{0.5cm}

\emph{published in Scholarpedia, 3(5):3930 --- accepted May 20, 2008}
 
\end{center}

\vspace{0.2cm}

\begin{center}
{\small \bf {ABSTRACT}}
\end{center}

Gravitational \emph{N-body simulations}, that is numerical solutions
of the equations of motions for N particles interacting
gravitationally, are widely used tools in astrophysics, with
applications from few body or solar system like systems all the way up
to galactic and cosmological scales. In this article we present a
summary review of the field highlighting the main methods for N-body
simulations and the astrophysical context in which they are usually
applied.


\section{Introduction}

The underlying dynamics relevant in the astrophysical context for of a
system of N particles interacting gravitationally is typically
Newton's law plus, in case, an external potential field (see however
below for a discussion of N-body simulations in general
relativity). The force $\vec{F}_i$ acting on particle
$i$ of mass $m_i$ is:

\begin{equation}\label{eq:newton}
\vec{F}_i = - \sum_{j \ne i}  G \frac{m_i m_j (\vec{r}_i-\vec{r}_j)}{|\vec{r_i}-\vec{r_j}|^3 } - \vec{\nabla} \cdot \phi_{ext}(\vec{r}_i),  
\end{equation}

\noindent where $G=6.67300 \cdot 10^{-11}$ $m^{3}$
$kg^{-1}$ $s^{-2}$ is the gravitational
constant, and $\phi_{ext}$ is the external potential. The problem
is thus a set of non-linear second order ordinary differential
equations relating the acceleration $\partial^2 \vec{r_i} /
\partial t^2 = \vec{F}_i /m_i$ with the position of all the
particles in the system.

Once a set of initial condition is specified (for example the initial
positions $\vec{r}_i$ and velocities $\vec{v}_i \equiv
\partial \vec{r}_i / \partial t $ of all particles) it exists a
unique solution, analytical only for up to two bodies, while larger N
require numerical integration (e.g. see Press et al. 2007). However
special care must be employed to ensure both accuracy and
efficiency. In fact, the gravitational force
(eq. \ref{eq:newton}) presents a singularity when the distance of
two particles approaches 0, which can lead to arbitrarily large
relative velocities. In addition, given the non-linear nature of the
equations, the singularities are movable, that is they depend on the
specific choice of initial conditions. In contrast, all singularities
in linear ordinary differential equations are independent of initial
conditions and thus easier to treat. Therefore constant timestep
methods are unable to guarantee a given accuracy in the
case of gravitational dynamics and lead to unphysical accelerations
during close encounters, which in turn may create unbound stars. A
shared adaptive timestep scheme can correctly follow a close
encounter, but the price is paid in terms of efficiency as all the
other particles of the system are evolved on the timescale of the
encounter, which may be several orders of magnitude smaller than the
global timescale, resulting essentially in a freezing of the
system.

The singularity may be avoided by introducing a smoothing length in
Eq.~\ref{eq:newton} (e.g. see Aarseth 1963), that is by modifying the
gravitational interaction at small scales. For example:
\begin{equation}  
  \vec{F}_i = - \sum_{j \neq i} \frac{G m_i m_j
    (\vec{r}_i - \vec{r}_j)}{(|\vec{r}_i - \vec{r}_j|^2 + \epsilon^2)^{3/2}} ,
\end{equation}
\noindent where $\epsilon > 0$ is the softening, or smoothing length, that is a
typical distance below which the gravitational interaction is
suppressed. To minimize the force errors and the global impact of the
softening for distances larger than $\epsilon$, finite size
kernels that ensure continuous derivatives of the force may be
employed (e.g., see Dehnen 2001). This strategy effectively
suppresses binary formation and strong gravitational interactions, but
at the price of altering the dynamics of the system.

The computational complexity of the numerical solution of a N-body
system for a fixed number of timesteps scales as $N^2$, as the
evaluation of the force on each particle requires to take into account
contributions from all other members of the system. For example,
considering a single state of the art cpu core (speed $\approx 5$
GFlops), a single force evaluation through a direct method would
require about 1 second for a system with $N=10^4$ particles (assuming
10 floating point operations per pair of particles) and more than a
week for $N=10^7$.

The arbitrarily large dynamic range in the unsoftened dynamics and the
expensive evaluation of the force have led to the development of a
wide number of numerical techniques aimed at obtaining a reliable
numerical solution with the minimum amount of computational resources,
depending on the astrophysical problem of interest. Here we start by
discussing the different astrophysical contexts where N-body
simulations are routinely employed and we then present the state of
the art techniques for these problems.

\section{Astrophysical domains and timescales}

N-body simulations are applied to a wide range of different
astrophysical problems so that the most appropriate technique to use
depends on the specific context, and in particular on the timescale
and collisionality of the problem.  

\subsection{Timescales, Equilibrium and Collisionality}
 
A system of N particles interacting gravitationally with total mass M
and a reference dimension R (for example the radius containing half of
the total mass) reaches a dynamic equilibrium state on a timescale
comparable to a few times the typical time ($T_{cr}$) needed for a
particle to cross the system ($T_{cr} \approx 1/\sqrt{GM/R^3}$). This
is the response time needed to settle down to virial equilibrium, that
is $2K/|W|=1$, where $K$ is the kinetic energy of the system: $K=1/2
\sum_{i=1,N} m_i |\vec{v}_i|^2 $, and W is its potential energy: $W =
- 1/2 \sum_{i \ne j} G m_i m_j /|\vec{r}_i-\vec{r}_j|$ (assuming no
external field). If the system is initially out of equilibrium, this is
reached through mixing in phase space due to fluctuations of the
gravitational potential, a process called violent relaxation
(Lynden-Bell 1967).
 
Once the system is in dynamic equilibrium a long term evolution is
possible, driven by two-body relaxation. Energy is slowly exchanged
between particles and the system tends to evolve toward thermodynamic
equilibrium and energy equipartition. The timescale ($T_{rel}$) for
this process depends on the number of particles and on the geometry of
the system: $T_{rel} \propto N/log(0.11 N) T_{cr}$ (e.g. see Spitzer
1987). N-body systems such as galaxies and dark matter halos have a
relaxation time much longer than the life of the Universe and are thus
considered collisionless systems. Smaller systems, such as globular
and open clusters, are instead collisional, as the relaxation time is
shorter than their age. Two body relaxation is also suppressed when
one particle in the system dominates the gravitational potential, such
as in the case of solar system dynamics, where planets are essentially
quasi-test particles.

Close encounters between three or more particles not only contribute
to energy exchange, but can also lead to the formation of bound
subsystems (mainly binaries). The formation and evolution of a binary
population is best followed through direct, unsoftened, N-body
techniques.

A self-gravitating N-body system made of single particles has a
negative specific heat, that is it increases its kinetic energy as a
result of energy losses (Lynden-Bell \& Wood 1968). This is a consequence of
the virial theorem and qualitatively it is analogous to the
acceleration of a Earth artificial satellite in presence of
atmospheric drag. A negative specific heat system is thermodynamically unstable and over
the two body relaxation timescale it evolves toward a gravothermal
collapse, creating a core-halo structure, where the core progressively
increases its concentration, fueling an overall halo expansion. The
collapse is eventually halted once three body interactions lead to the
formation of binaries. The so called "core collapsed globular
clusters" are considered to be formed as a result of this mechanism.

\subsection{Mean field approach: the Boltzmann equation}

A system of N particles interacting gravitationally defines a 6N+1
dimensional phase space given by the N position and velocity vectors
associated to each particle at each time t. The solution of the N-body
problem defines a trajectory in this phase space. If the number of
particles is large enough, that is if the two body relaxation time is
long compared to the time-frame one is interested in, then a
statistical description of the problem is possible. This allows us to pass from a 6N+1 to a 6+1 dimension
phase space. The idea is to construct a mean field
description of the dynamical system in terms of a single particle
distribution function $f(\vec{x},\vec{v},t) $, where
$f(\vec{x},\vec{v},t) d^3x d^3v $ is proportional to the
probability of finding a particle in a 6D element of volume $
d^3x d^3v $ centered around position $\vec{r}$ and
velocity $\vec{v}$ at time t. Within this simplified framework the
knowledge of the distribution function uniquely defines all the
properties of the system. The dynamic is described by the
collisionless Boltzmann equation, which derives essentially from the
Liouville theorem:
\begin{equation}
\frac{D f}{D t} = \frac{\partial f}{\partial t} + \vec{v} \cdot \frac{\partial f}{\partial \vec{x}} - \frac{\partial \phi_T}{\partial \vec{x}} \cdot \frac{\partial f}{\partial \vec{v}}= 0,
\end{equation}
\noindent where the total potential field $\phi_T = \phi_{ext}(\vec{x},t)+
\phi(\vec{x},t)$ is the sum of an external potential plus the
self-consistent field $\phi(\vec{x},t)$ defined from the
distribution function itself through the solution of the Poisson
equation:
\begin{equation}
\nabla^2 \phi(\vec{x},t) = 4 \pi G  \rho(\vec{r},t), 
\end{equation}
\noindent where $ \rho(\vec{r},t) = \int f(\vec{x},\vec{v},t) d^3v $.

Given its high dimensionality (6+1), the collisionless Boltzmann
equation is usually solved by sampling the initial distribution
function and then by evolving the resulting N-body system by means of
a numerical method that suppresses two body interactions at small
scales. The interaction is softened not only for computational
convenience to limit the maximum relative velocity during close
encounters but especially to prevent artificial formation of
binaries. This is because a simulation particle in a collisionless run
represents in reality an ensemble of real particles (e.g. galaxies
contain $10^{11}$ stars but simulations typically use only $N \in
[10^6:10^9]$). Note however that two body relaxation is driven by
close as well as by distant encounters, so softening does not suppress
it. In principle any numerical method that has a small scale softening
is appropriate for following collisionless dynamics.

A mean field description for an N-body system is possible also for
collisional systems, that is when the relaxation time is comparable to
or shorter than the timeframe of interest. In this case the
collisionless Boltzmann equation is modified by the introduction of a
collision operator $C[f]$ on its right side:

\begin{equation}
\frac{D f}{D t} = \frac{\partial f}{\partial t} + \vec{v} \cdot \frac{\partial f}{\partial \vec{x}} - \frac{\partial \phi_T}{\partial \vec{x}} \cdot \frac{\partial f}{\partial \vec{v}}= C[f].
\end{equation}

In this framework the operator $C[f]$ describes the
probability for particles to enter/leave a phase space element as a
result of gravitational encounters. The collision operator
C is generally constructed assuming that encounters are:
\begin{enumerate}
\item Markov processes, that is C depends only on the present state of
the system;
\item local, that is only the velocity of the particles are
changed and not their positions;
\item weak, that is the typical velocity change is much smaller than
the velocity itself. 
\end{enumerate}
\noindent Under these assumptions Monte Carlo methods are available to
solve the dynamics of the system (see next section). Applications of
the collision operator include dynamics of globular clusters and of
self-interacting dark matter.

\subsection{Mean Field Approach: analogies and differences with fluid dynamics}

The velocity moments of the Boltzmann Equation define a set of
equations known as the Jeans Equations (e.g. Binney \& Tremaine
2007). The first three equations of the set are formally identical to
the Navier-Stokes equations for a self-gravitating gas and, like in
the fluid-dynamics analogy, express the conservation of mass, momentum
and energy. Therefore the numerical algorithms developed to follow the
dynamics of N-body systems find a wide application also in the context
of fluid-dynamics, with one important example being the Smoothed
Particle Hydrodynamics (SPH) method (Gingold \& Monaghan 1977). The
fundamental difference between the two cases is that the Jeans
equations are derived in the limit of a collisionless system, while
the Navier-Stokes equations assume a highly collisional system, with
the mean free path of a particle approaching zero. For fluids, this
leads to the definition of an equation of state, which closes the
Navier-Stokes equations. The Jeans Equations are instead an infinite
open set, where the {\it{n-th}} velocity moment depends on the {\it{n-th+1}} moment.

\subsection{Astrophysical domains}

Based on the previous considerations about collisionality and
timescales, four main astrophysical domains for N-body simulations can
be identified, each requiring a different numerical technique to
guarantee maximum performance and accuracy:

\emph{Celestial mechanics} (solar and extrasolar planetary systems). Here
a single body dominates the gravitational field and all the other
objects move almost like test particles, subject to reciprocal
perturbations. In this framework very high accuracy is required to
correctly evaluate the perturbative terms and to avoid being dominated
by numerical noise such as time discretization and round-offs errors.

\emph{Dense stellar systems}, such as open clusters and globular
clusters. These collisional systems made of components of roughly
equal mass present a rich dynamics, with multiple close encounters of
stars. The evolution requires to be followed on a relaxation timescale
with a correct description of the short range interactions.

\emph{Sphere of influence of a massive BH} at the center of a stellar
system. The sphere of influence of a BH is the volume within which the
gravity of the BH dominates over that of the other particles. The
situation resembles that of solar system dynamics, but here given the
very high density of stars two body encounters are frequent, making
the problem a difficult hybrid between the two previous cases. In
addition, Post Newtonian physics may need to be included if high
accuracy is required in the proximity of the BH.

\emph{Galaxy dynamics and cosmology}. Galaxies, and especially dark
matter halos, are constituted by a very large number of particles, so
that their dynamics can be well described in terms of a mean
field. Close encounters are not important and softening is usually
employed in these N-body simulations to avoid the unphysical formation
of binaries. Within this class, Self-Interacting Dark Matter Particles
need a special mention: if dark matter halos are made of Weakly
Interacting Massive Particles, then their dynamics can be modified by
non-gravitational self-interactions, especially effective at the
center of cuspy dark halos. The dynamics of such a system is described
by the Collisional Boltzmann Equation, which can be approximately
solved using Fokker-Plank methods.

\section{Newtonian gravity: methods}

The history of N-body simulations starts with a pioneering attempt by
Holmberg (1941), who followed the evolution of a 37 particle system,
where the force was calculated using lightbulbs and galvanometers
(taking advantage of the same $r^{-2}$ scaling of electromagnetic and
gravitational interactions). Computer simulations started in the early
sixties using up to 100 particles (e.g. see von Hoerner 1960 and
Aarseth 1963) and had their full bloom in the eighties with the
development of fast and efficient algorithms to deal with
collisionless systems, such as particle-mesh codes (see Hockney \&
Eastwood 1988 and references therein) and the tree method (Barnes \&
Hut 1986). At the same time regularization techniques were developed
to deal with close encounters and binary dynamics in the case of
direct simulations of a collisional system (e.g. see Aarseth's NBODY-X
code series based on KS and chain regularization - Aarseth 2003 and
references therein). These algorithm advancements were coupled with
tremendous progresses in the hardware, with the cpu speed growing
exponentially. In addition to parallelization of serial codes, the
field advanced also thanks to special purpose hardware, such as the
GRAPE (Makino et al. 1997). Today's (2008) N-body simulations are
performed with up to $N=10^5$ (e.g. see Baumgardt \& Makino 2003) for
direct integration codes over a two-body relaxation timescale and up
to $N=10^{10}$ for collisionless dynamics/cosmology (e.g. see the
Millennium Run - Springel et al. 2005). In the context of planetary
dynamics, self-gravitating systems of disk/ring particles with
$N\approx 10^6 $ can be followed over hundreds of dynamical times
(e.g. Richardson et al. 2000). Major breakthroughs are also expected
in the near future thanks both to the next generation GRAPE-DR and to
double precision graphic processing units, which provide extremely
cost competitive high performance computing capabilities.

\subsection{Direct methods}

Direct methods do not introduce approximations in the solution of the
equations of motions and thus deliver the highest accuracy at the
price of the longest computation time, of order $O(N^2)$ per
timestep. Integration is performed using adaptive (individual)
timesteps and commonly a fourth order Hermite integrator. Close
encounters and bound subsystems are treated exactly in terms of
Kustaanheimo-Steifel transformations.  These essentially consist in
transformations of coordinates using a perturbative approach over the
analytical two body solution. If more than two particles have a strong
mutual interaction, then a chain regularization strategy (Mikkola
1990) can be used, which consists in recasting the problem in terms of
a series of separate Kustaanheimo-Steifel interactions. A state of the
art, publicly available, serial direct N-body integrator is Aarseth's
NBODY6.  Even with this specialized software, the number of particles
that can be effectively followed for timescales comparable to the
Hubble time is limited. For example, if one is interested in the
dynamical evolution of globular clusters, currently about $N=20000$ is
the practical limit for a serial run, as such a run takes about $1000$
cpu hours. A run with $10^6$ particles carried out for a similar
number of relaxation times $T_{rel}$ would require about $10^8$ cpu
hours. The algorithm can be parallelized, but in practice load
imbalances may saturate the gain in efficiency, so some of the most
cpu demanding simulations have been carried out on special purpose
hardware, such as the GRAPE, where the chip architecture has been
optimized to compute gravitational interactions, thus delivering
Teraflops performance.

\subsection{Tree codes}

The tree code method (Barnes \& Hut 1986) provides a fast, general
integrator for collisionless systems, when close encounters are not
important and where the force contributions from very distant
particles does not need to be computed at very high accuracy. In fact,
with a tree code, small scale, strong interactions are typically
softened (but see McMillan \& Aarseth 1993), while the potentials due
to distant groups of particles are approximated by multipole
expansions about the group centers of mass. The resulting computation
time that scale as $O(N log(N))$ but the approximations introduce some
(small) errors. The errors in the long-range component of the
gravitational acceleration are controlled by a single parameter (the
so called opening angle) that determines how small and distant a group
of particles must be to use the approximation. This strategy works
well to keep the average force error low, but a worst case scenario
analysis highlights that unbound errors can arise for rare, but
astrophysically reasonable configurations, such as that of the classic
"exploding galaxy" (Salmon \& Warren 1994). In addition, force errors
from the tree code may lead to violation of momentum
conservation. Typical implementations of the tree code expand the
potentials to quadrupole order and construct a tree hierarchy of
particles using a recursive binary splitting algorithm. The tree does
not need to be recomputed from scratch at every timestep, saving
significant cpu time. Systems with several hundred thousands of
collisionless particles can be easily simulated on a GFlops
workstation for a Hubble time using this method.

\subsection{Fast Multipole Methods}

A standard tree code implementation does not take advantage of the
fact that nearby particles will be subject to a similar acceleration
due to distant groups of particles. The Fast Multipole Method
(Greengard \& Rokhlin 1987) exploit this idea and uses a multipole
expansion to compute the force from a distant source cell within a
sink cell. This additional approximation of the gravitational
interaction was claimed to reduce the complexity from $O(N log(N))$ to
$O(N)$, but the exact scaling seems implementation dependent and has
been debated in the literature (e.g. see Dehnen 2000 and references
therein). One advantage of the fast multipole method is that the
symmetry in the treatment of sink and source cells with respect to the
multipole expansion can guarantee an exact conservation of the
momentum. Recent successful implementations of fast multipole codes or
hybrids with tree code scheme, include Dehnen's Cartesian expansion
scheme (the GyrfalcON code- Dehnen 2000) and PKDGRAV (Stadel 2001).

\subsection{Particle-mesh codes}

The particle mesh method represents another route to speed up direct
force evaluation for collisionless systems. In this case the
gravitational potential of the system is constructed over a grid
starting from the density field and by solving the associated Poisson
equation. Particles do not interact directly between each other but
only through a mean field. The method essentially softens the
gravitational interactions at small scales, that is below the cell
length. The density field is constructed using a kernel to split the
mass of the particles to the grid cells around the particle
position. The simplest choice is to assign all the mass to a single
cell, but this leads to significant force fluctuations, which can be
reduced using a cloud in cell (8 points) or a triangular shaped cloud
(27 points) kernel. The Poisson equation is typically solved using a Fast
Fourier Transform, but other grid methods such as successive overrelaxation can also be used - e.g. see Bodenheimer et al. (2007). The deriving force, defined on the grid, is then
assigned back to the particles using the same kernel employed for the
density field construction, in order to avoid spurious self
forces. The method achieves a linear complexity in the number of
particles and ($O(N_g log(N_g)$) in the number of grid
cells (this latter scaling is that of the FFT method). The price to
pay is in terms of short range accuracy as the force is a poor
approximation of Newton's law up to several grid spacing of
distance.

\subsection{Adaptive Mesh Refinement method}

The dynamic range of particle-mesh codes can be increased by using an
adaptive rather than a static grid to solve the Poisson Equation. In
the Adaptive Mesh Refinement (AMR) method the grid elements are
concentrated where a higher resolution is needed, for example around
the highest density regions. One possibility to obtain an adaptive
resolution is to first construct a low-resolution solution of the
Poisson Equation and then to progressively refine regions where the
local truncation error (estimated through the Richardson
extrapolation) is highest. A multigrid structure needs to take into
account issues such as matching the solution at the grid interfaces.
AMR codes are well suited for cosmological simulations (e.g. see the ENZO code, Bryan \& Norman
1998).

\subsection{Self consistent field methods}

A variant over the Particle Mesh code is the expansion of the density
and potential of the system in terms of a basis of orthogonal
eigenfunctions. Clutton-Brock (1972) was one of the first to apply
this idea in stellar dynamics, while a modern implementation is that
of Hernquist \& Ostriker (1992). This method guarantees at fixed
computational resources a higher accuracy than the tree code and the
particle mesh algorithms, provided that the set of basis function is
appropriately selected. This limits in practice a general application
of the method, which remains however very competitive for the study of
the dynamical stability of collisionless systems constructed from
distributions functions models.

\subsection{P3M and PM-Tree codes}

In order to increase the force resolution of particle mesh codes it
has been proposed to couple a mean field description on large scales
with a direct, softened, treatment of the gravitational interactions
on distances of the order of or below a few grid spacing. This method
is called $P^3M$ (Hockney \& Eastwood 1988):
Particle-Particle-Particle-Mesh and efficiently increases the dynamic
range of the parent PM algorithm. However in presence of strong
clustering a large number of particles will interact directly between
each other, slowing down significantly the computation to
$O(N^2)$. This problem can be resolved by using adaptive meshes, so
that the spatial resolution is refined in regions of high
density. Adaptive $P^3M$ codes have a computational cost which scales
as $O(N log(N))$, like in a tree code. Finally another possibility is
to resort to a tree code for the short range force evaluation leading
to a hybrid PM-Tree scheme. These methods are generally extremely well
suited for cosmological simulations, for example see Gadget2 (Springel
2005).

\subsection{Celestial mechanics codes}

Computational Celestial Mechanics refers to a series of methods
targeted at studying the dynamics of small N systems ($ N \lesssim 20
$). The smallest non trivial N is N=3, that is the three body problem,
which has many applications ranging from space flight to planets
satellite motions and to binary-single stars encounters. Celestial
mechanics requires extremely high precision given the chaotic nature
of the N-body problem. Numerical methods are based on the use of local
system of coordinates, to fight round-off errors in systems with a
wide dynamic range, such as in the study of star-planet-satellite
problems, as well as on the variational equations formalism and on
perturbation theory to take advantage of the analytical, unperturbed
motion of planets in the gravitation field of their star (e.g. see
Beutler 2005). In this context symplectic integrators are widely used
(e.g. see Wisdom \& Holman 1991; Leimkuhler \& Reich 2005).

\section{Mean Field Methods}

As an alternative to particle based N-body methods, the dynamics of a
system of particles interacting gravitationally can be followed by
solving the time dependent Boltzmann Equation coupled with the
self-consistent Poisson equation.

\subsection{Grid based solvers for the Collisionless Boltzmann Equation}

This approach can take advantage of standard computational methods
developed to solve partial differential equations, such as successive
over-relaxation and conjugate gradient methods. However it requires to
solve a highly dimensional (6D+time) non-linear system of partial
differential equations. In general, the bottleneck is thus the very
large amount of memory needed (for example, Terabites just to have a
moderate resolution grid with 100 elements in each dimension). However
this method is competitive if the astrophysical problem of interest
presents symmetries that reduce the number of dimensions needed in the
model. For example, in the case of globular cluster dynamics a very
good approximation can be obtained via a 3 dimensional model by
assuming spherical symmetry in the position space (1D) and radial
anisotropy in the velocity space (2D).

\subsection{Fokker-Planck and Monte Carlo methods}

These methods solve the collisional Boltzmann equation starting from a
given distribution function and by following test particles in the six
dimensional position-velocity phase space. At each timestep the
velocity of the particles is perturbed by random fluctuations
accordingly to the assumed form for the collision operator $ C[f]$,
which depends on computed cross sections for two, three and four body
encounters. The complexity of Monte Carlo codes is linear with the
number of particles and thus a realistic number of particles can be
used for simulations of collisional systems with $ N>10^5 $ with a
serial code. The method is ideal for exploring grids of initial
conditions, after proper validation through comparison with direct
integration (e.g. see Heggie et al. 2006).

\subsection{Beyond Newton: strong gravitational fields}

In presence of a strong gravitational field, such as that in the
proximity of the event horizon of a black hole, N-body simulations
cannot be based on Newtonian physics, but must take into account a
general relativity framework. As a numerical solution of the Einstein
equation is extremely challenging, Post-Newtonian approximations are
used when the gravitational field does not deviate too much from the
Newtonian case. Post-Newtonian corrections are typically good enough
to treat most astrophysical problems of the dynamics of stars around a
black hole. A full general relativity framework is only required to
study the merging and gravitational waves emission of two black-holes
(e.g. see Baker et al. 2006).

\section{Hardware}

An alternative approach to increase the efficiency of numerical
solution of the N-body problem is to optimize the hardware. For direct
simulations this approach can be very effective, thanks to the fact
that the bottle neck of computation is just the evaluation of the
gravitational force, which has a very simple expression. Along this
route the GRAPE (GRavityPipE) concept has been extremely
effective. The basic idea is to optimize a hardware pipeline to
compute $(\vec{r_i}-\vec{r_j})/| \vec{r_i}-\vec{r_j}|^3 $. This special purpose hardware can then be interfaced with a
general purpose computer, which takes care of all the other numerical
operations required to solve the equations of motions. With the
GRAPE-6, the largest simulation on a collisional timescale published
to date has N=131028 (Baumgardt \& Makino 2003). 

Another recent promising hardware development is the possibility to
use Graphic Cards (GPUs) to carry out the cpu intensive force
evaluation. The performance of current generation of GPUs appears to
be superior (in terms of Flops/\$ ratio) to that of the GRAPE6 series
(Portegies-Zwart et al. 2007) even if one important limitation of GPUs
is that they currently operate in single precision.

\section{Simulation environments}

In addition to the availability of stand-alone codes, several software
environments have been created that contain various tools to set up
initial conditions, run simulations, and analyze and visualize their
results.  Some examples are NEMO, Starlab, ACS and MUSE (see below for
links to their web-pages).

\section{Suggested readings}

\subsection{Books}

\begin{itemize}
\item "Computer Simulation Using Particles" Hockney, R.W. and Eastwood, J.W. 1988 
\item "Gravitational N-Body Simulations: Tools and Algorithms" Aarseth, S. 2003
\item "The Gravitational MillionBody Problem" Heggie, D.C. and Hut, P. 2003
\item "Methods of Celestial Mechanics" Beutler, G. 2005
\item "Numerical Recipes" Press, W.H., Teukolsky, S.A., Vetterling, W.T. and Flannery B.P. 2007
\item "Numerical Methods in Astrophysics: An Introduction" Bodenheimer, P., Laughlin, G.P., Rozyczka, M. and Yorke, H.W. 2007
\end{itemize}

\subsection{Review articles}

* "Simulations of Structure Formation in the Universe" Bertschinger, E. 1998, ARA\&A, 36, 599

\subsection{Web Material}

\begin{itemize}
\item "The N-body Constitution" by Lake, G, Katz, N., Quinn T. and
Stadel. J. (http://www-hpcc.astro.washington.edu/old\_content/siamhtml/siamhtml.html)
\end{itemize}

\section{Open source codes}

\begin{itemize}

\item Aarseth's direct integration codes: http://www.ast.cam.ac.uk/\~sverre/web/pages/nbody.htm

\item ACS, a collection of tools and introductory texts: http://www.artcompsci.org/

\item ENZO, a cosmological AMR code: http://lca.ucsd.edu/portal/software/enzo

\item Gadget2, a cosmological PM-tree+SPH code (massively parallel): http://www.mpa-garching.mpg.de/gadget/

\item Mercury (a mixed variable symplectic integrator code for planetary dynamics): http://www.arm.ac.uk/\~jec/home.html

\item MUSE, a software framework for simulations of dense stellar systems: http://muse.li/

\item NEMO collection (includes particle-grid and tree codes): http://bima.astro.umd.edu/nemo/

\item Starlab (including the direct integration Kira code): http://www.ids.ias.edu/\~starlab/starlab.html

\end{itemize}

\section*{Acknowledgments}

We thank Douglas Heggie, Derek Richardson and two anonymous referees
for useful comments and suggestions. Further suggestions and comments
are very welcome as it is in the spirit of Scholarpedia to keep the
articles up-to-date.

\section*{References}

{\small{ 
\begin{enumerate}

\item Aarseth, S. 1963, MNRAS, 126, 223
\item Aarseth, S. 2003,  "Gravitational N-Body Simulations: Tools and Algorithms", Cambridge University Press
\item Baker, J.G. et al. 2006, ApJ, 653, 93 
\item Barnes, J.E. and Hut, P. 1986, Nature, 324, 466 
\item Baumgardt, H. and Makino, J. 2003, MNRAS, 340, 227
\item Beutler G. 2005, "Methods of Celestial Mechanics", Springer 
\item Binney J. \& Tremaine S.  1987, "Galactic Dynamics", Princeton University Press
\item Bodenheimer, P., Laughlin, G.P., Rozyczka, M. and Yorke, H.W. 2007, "Numerical Methods in Astrophysics: An Introduction", Taylor \& Francis
\item Bryan G.L. and Norman, M.L. 1998, ApJ, 495, 80 
\item Clutton-Brock, M. 1972, Ap\&SS, 16, 101
\item Dehnen, W. 2001, MNRAS, 324, 273
\item Dehnen, W. 2000, ApJL, 536, 39
\item Gingold, R.A. and Monaghan, J.J. 1977, MNRAS, 181, 375 
\item Greengard, L. \& Rokhlin, V. 1987, J. comput. Phys., 73, 325
\item Heggie, D.C., Trenti, M. and Hut, P. 2006, MNRAS, 368, 677
\item Hernquist, L and Barnes, J.E. 1990, ApJ, 349, 562
\item Hockney, R.W. and Eastwood, J.W. 1988, "Computer Simulation Using Particles", Taylor \& Francis
\item Holmberg, E. 1941, ApJ, 94, 385
\item Leimkuhler, B. and Sebastian R. 2005, "Simulating Hamiltonian Dynamics", Cambridge University Press 
\item Lynden-Bell, D. 1967, MNRAS, 136, 101
\item Lynden-Bell, D. and Wood, R. 1968, MNRAS, 138, 495
\item Makino, J., Fukushige, T., Koga, M. and Namura, K. 2003, PASJ, 55, 1163
\item Portegies-Zwart, S.F., Belleman, R.G. and Geldof, P.M. 2007, New Astronomy, 12, 641
\item Press, W.H., Teukolsky, S.A., Vetterling, W.T. and Flannery B.P. 2007, "Numerical Recipes", Cambridge University Press
\item Richardson, D.C., Quinn, T., Stadel, J. and Lake, G. 2000, Icarus, 143, 45
\item Salmon J.K. and Warren M.S. 1994, J. Comp. Phys., 111, 136.
\item Spitzer, L. 1987, "Dynamical Evolution of Globular Clusters", Princeton University Press
\item Springel, V. 2005, MNRAS, 364, 1105 
\item Springel, V. et al. 2005, Nature, 435, 629
\item Stadel J. 2001, PhD. Thesis, University of Washington
\item von Hoerner, S. 1960, Z. Astrophys. 50, 184
\item Wisdom, J. and Holman, M. 1991, AJ, 102, 1528

\end{enumerate}
}}

\end{document}